\documentclass[conference]{IEEEtran}
\IEEEoverridecommandlockouts
\usepackage{cite}
\usepackage{caption}
\usepackage{amsmath,amssymb,amsfonts}
\usepackage{algorithmic}
\usepackage{graphicx}
\usepackage{textcomp}
\usepackage{xcolor}
\usepackage[a4paper, total={184mm,239mm}]{geometry}
\def\BibTeX{{\rm B\kern-.05em{\sc i\kern-.025em b}\kern-.08em
    T\kern-.1667em\lower.7ex\hbox{E}\kern-.125emX}}
\begin{document}

\title{Reconfigurable Frequency Multipliers Based on Complementary Ferroelectric Transistors\\
}
 \author{\small{}
         Haotian~Xu$^1$, Jianyi~Yang$^1$, Cheng~Zhuo$^{1,2,*}$, Thomas~K{\"a}mpfe$^3$, Kai~Ni$^4$, and Xunzhao~Yin$^{1,2,*}$\\
    $^1$Zhejiang University, Hangzhou, China\\
    $^2$Key Laboratory of Collaborative Sensing and Autonomous Unmanned Systems of Zhejiang Province, Hangzhou, China\\
    $^3$Fraunhofer IPMS, Dresden, Germany; $^4$Department of Electrical Engineering, University of Notre Dame, Notre Dame, USA\\
    $^*$Corresponding authors, email: \{czhuo, xzyin1\}@zju.edu.cn
}

\maketitle

\begin{abstract}
Frequency multipliers, a class of essential electronic components, play a pivotal role in contemporary signal processing and communication systems. They serve as crucial building blocks for generating high-frequency signals by multiplying the frequency of an input signal. 
However, traditional frequency multipliers that rely on nonlinear devices often require energy- and area-consuming  filtering  and amplification circuits, 
and emerging designs based on an ambipolar ferroelectric transistor require costly non-trivial characteristic tuning or complex technology process.
In this paper, we show that a pair of standard ferroelectric field effect transistors (FeFETs) can be used to build  compact frequency multipliers without aforementioned technology issues.  
By leveraging the tunable parabolic shape of the 2FeFET structures' transfer characteristics, we propose four  reconfigurable  frequency multipliers, which can switch between signal transmission and frequency doubling.
Furthermore, based on the  2FeFET structures, we propose  four frequency multipliers that realize triple, quadruple frequency modes, elucidating a scalable methodology to generate more multiplication  harmonics of the input frequency.
Performance metrics such as maximum operating frequency, power, etc., are evaluated and compared with existing works.
We also implement a practical case of frequency modulation scheme based on the proposed reconfigurable multipliers without additional devices.
Our work provides a novel path of scalable and reconfigurable frequency multiplier designs based on devices that have characteristics similar to FeFETs, and show that FeFETs are a promising candidate for signal processing and communication systems in terms of maximum  operating frequency and power.

\end{abstract}

\vspace{-1ex}
\section{Introduction}
\label{sec:intro}
The demand for higher frequency signals in modern electronics has necessitated the development of innovative techniques for frequency generation \cite{lanza2022memristive, 6830102,zheng2023coarray, abdulazhanov2023reconfigurable, maestrini2008terahertz}. 
Frequency multipliers, a fundamental category of electronic circuits, have emerged as indispensable tools in this endeavor \cite{890315}. 
A frequency multiplier 
produces an output signal with a frequency that is an integer multiple of the input frequency.
Most of this fundamental operation is achieved through nonlinear signal processing techniques, wherein the input signal is subjected to a series of transformations to generate a higher frequency output\cite{554612}. 


The currently widely used nonlinear devices include two types: Schottky diodes and metal oxide field effect transistors (MOSFETs). 
Schottky diodes are utilized as nonlinear elements in the frequency multipliers, which perform 
exceptionally well at high frequencies. 
However, the structural characteristics of Schottky diodes  pose challenges when integrating them with 
other components like amplifiers and oscillators. Additionally, Schottky diode based multipliers require significant input driving power to ensure sufficient conversion gains due to the inherent passivity of diodes \cite{4435094,4266491,6665892}.
Alternatively, MOSFET based multipliers are compatible with CMOS technology, and can leverage the transconductance characteristics of transistors to amplify voltage fluctuations at input and/or output, thereby enhancing the conversion gain even  with low input power \cite{Single-chip,3,4,5,6,7,ref1}.
However, such nonlinear devices not only generate the desired frequency signal but also produce extra harmonics, thus necessitating complex filtering circuits, which are often energy- and area-consuming \cite{4266491,ref1,7}. 
To date, 
more efficient frequency multiplication techniques are still undergoing.
 
Recently, 
innovative approaches harnessing the ambipolar parabolic characteristic curve of ferroelectric field effect transistors (FeFETs) to induce second harmonic  were introduced \cite{mulaosmanovic2020reconfigurable, xu2023challenges, zhu2023reconfigurable}. Through the controlled programming of  ambipolar FeFETs, the developed single-transistor frequency multipliers that can switch between signal transmission and frequency doubling have been demonstrated, exhibiting low power and compact area.
However, it is worth noting that 
these frequency multiplier are limited to operating at approximately 1 MHz/10 kHz and  achieving only first and second harmonic generation, which cannot meet the high-frequency requirements of  communication systems. Furthermore, they suffer from either the non-trivial  tuning that ensures the symmetrical ambipolarity, or costly and complex nanowire tunneling FET technology. 
More general reconfigurable frequency multiplier design with scalable and efficient harmonic generations  are highly desired.


In this paper, we propose  a   general and economic design methodology for building compact and reconfigurable frequency multipliers with scalable harmonic generation capability.
Taking the complementary FeFETs as  representative devices, we show how to build four 2FeFET structures (i.e., 2 n-type FeFETs (nFeFETs), 2 p-type FeFETs (pFeFETs),  nFeFET and pFeFET in parallel, nFeFET and pFeFET in series) that exhibit tunable parabolic shape of transfer characteristics.
Utilizing the 2FeFET structures, we  propose
four reconfigurable frequency multipliers that can switch between signal transmission and frequency doubling. 
To  accommodate the high-frequency requirements, we further propose four multipliers based on the 2FeFET structures, realizing the reconfigurability between various multiplication modes, e.g., first, second, third and fourth harmonics.
These reconfigurable frequency multipliers  are evaluated in terms of performance, power, area, etc., and a practical case of frequency shit keying (FSK) based on the multipliers is demonstrated.
The goal of this work is to provide guidance in the general and efficient frequency multiplier designs of FeFETs for their signal processing applications.
\vspace{-1ex}
\section{Background}
\label{sec:background}
In this section, we first introduce the complementary FeFET devices, then review existing frequency multiplier designs.

\begin{figure}[t]
\centerline{\includegraphics[width=8cm]{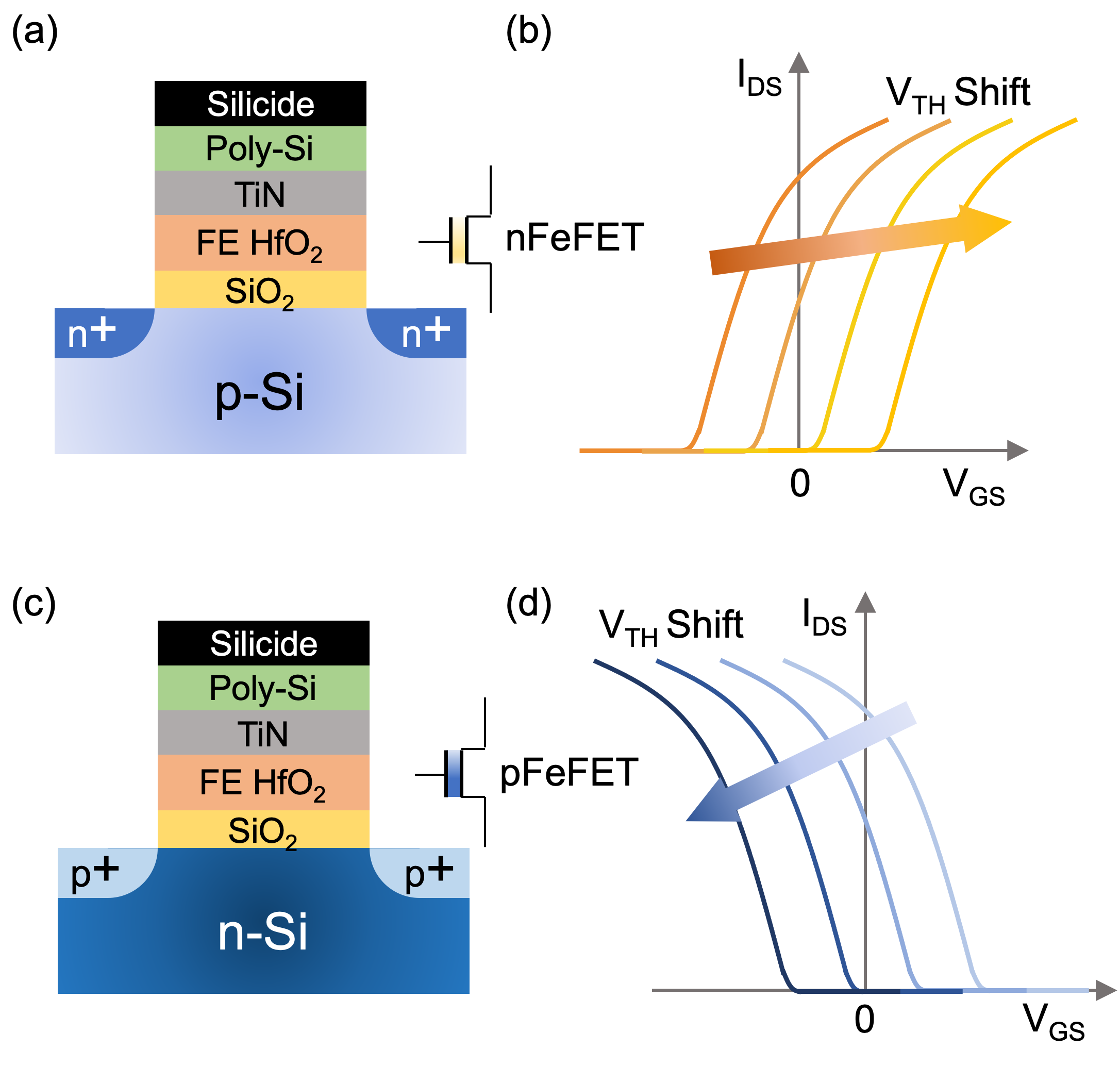}}
\vspace{-1ex}
\caption{
(a)/(b) N-type FeFET (nFeFET) and its $V_{TH}$ programmable characteristic curve  
(c)/(d)  P-type FeFET (pFeFET) and its  $V_{TH}$ programmable characteristics. 
}
\label{fig1}
\vspace{-2ex}
\end{figure}
\vspace{-1ex}
\subsection{FeFET basics}
\label{sec:fefet}
\vspace{-1ex}
FeFETs are fabricated by integrating a ferroelectric layer at the gate stack of a MOSFET, as shown in Fig. \ref{fig1}(a)/(c).
When a program voltage is applied to the gate, 
the degree and direction of polarization state of ferroelectric layer are set, 
making FeFETs non-volatile memory devices \cite{hu2021memory, Soliman_2020, yin2023ultracompact, huang2023fefet}. 
The varying polarization states corresponding to different program voltage pulses applied at  gate-source alter the threshold voltage ($V_{TH}$) of the FeFETs 
per Fig.\ref{fig1}(b),(d). 
It is through this characteristic that we achieve reconfigurable frequency multiplier designs.
The characteristic curves of nFeFETs and pFeFETs resemble those of nMOS and pMOS transistors \cite{pfefet}. 
Note that nFeFETs and pFeFETs exhibit symmetrical characteristics, a crucial aspect in  
complementary FeFET frequency multipliers designs.

\vspace{-1ex}
\subsection{Existing frequency multipliers}
\label{sec:relatedwork}
\vspace{-1ex}
In \cite{6665892}, a planar Schottky diode based  frequency doubler  is proposed, offering an impressive maximum operating frequency of 220GHz. 
However, Schottky diode multipliers requires exceptionally high input signal power, i.e., 200mW, to achieve high multiplication efficiency  due to the inherent passivity of diodes.
Another type of conventional high-power frequency doubler based on MOSFET can achieve 26GHz output at the cost of a relatively large chip size (0.28$mm^2$) and power consumption (10mW) \cite{ref1}. 
\begin{figure}[t]
\centerline{\includegraphics[width=1.0\columnwidth]{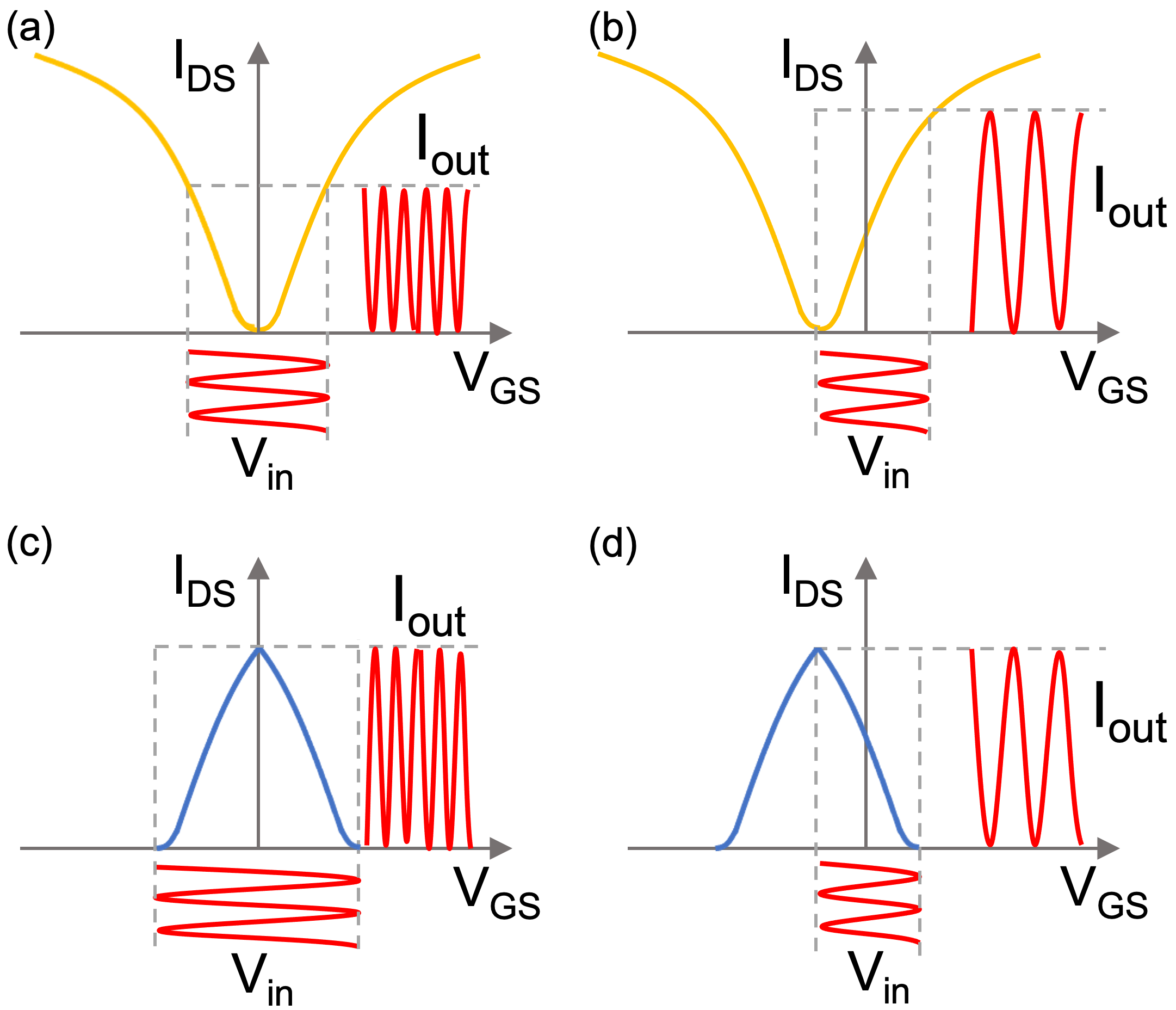}}
\vspace{-1ex}
\caption{
Principles of  frequency doubling and signal transmission utilizing (a)/(b) upward-opening or (c)/(d) downward-opening parabolic-shaped characteristic.
}
\label{fig2}
\vspace{-2ex}
\end{figure}
\begin{figure}[t]
\centerline{\includegraphics[width=1.0\columnwidth]{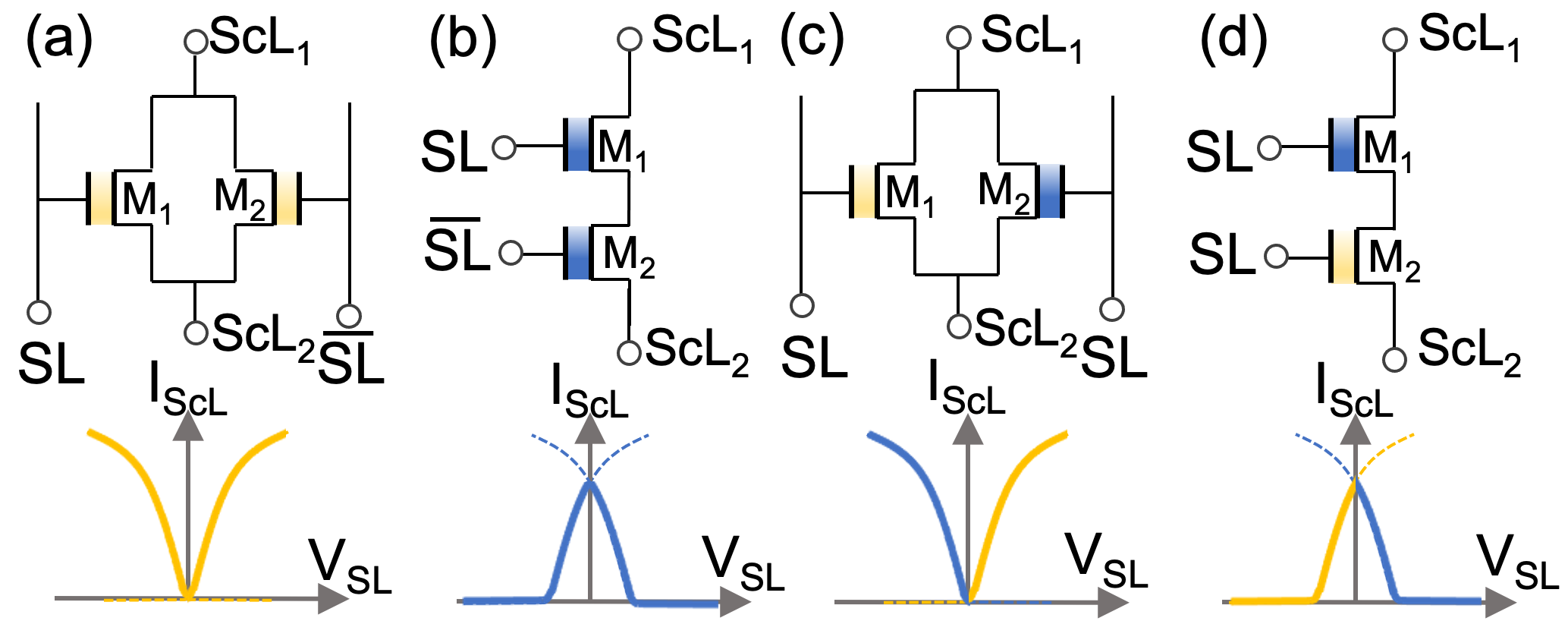}}
\vspace{-1ex}
\caption{
Four 2FeFET structures realizing parabolic-shape characteristics: 
(a) 2 nFeFETs in parallel for upward parabola.
(b) 2 pFeFETs in series for downward parabola.
(c) Complementary FeFETs in parallel for upward parabola.
(d) Complementary FeFETs in series for downward parabola.
}
\label{fig3}
 \vspace{-4ex}
\end{figure}
\begin{figure*}[t]
\centerline{\includegraphics[width=0.85\textwidth]{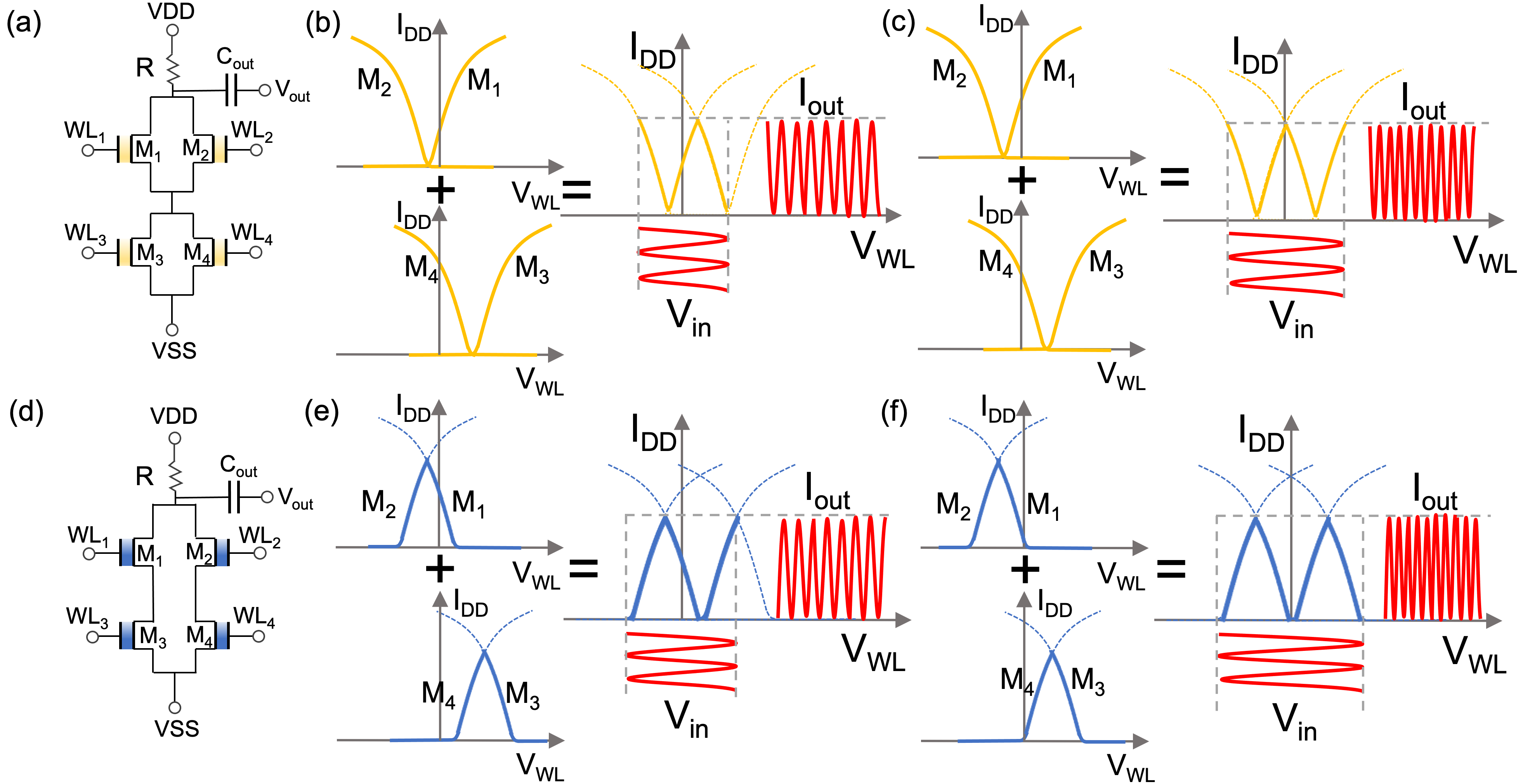}}
\vspace{-1ex}
\caption{
(a) 4n-serial design. (b)/(c) Leveraging the superimposed double-valley shape curve  to  generate triple/quadruple frequency.
(d) 4p-parallel design. (e)/(f) Leveraging the superimposed double-peak shape curve to generate triple/quadruple frequency.
}
\label{fig4}
\vspace{-4ex}
\end{figure*}
Emerging frequency multiplier designs harnessing the ambipolar parabolic characteristics of FeFET have been developed.
In \cite{mulaosmanovic2020reconfigurable}, modulation of a single FeFET's ambipolar current was achieved by altering the drain ($V_D$) and/or gate voltage ($V_G$), thereby approximating  transfer characteristic of the  FeFET to an ideal parabolic curve, thus enabling  frequency doubling and signal transmission, respectively. The conceptual  principles are illustrated in Fig. \ref{fig2}(a)/(b). 
However, in practice,  the modulation of symmetrical ambipolar characteristic of the single FeFET requires costly tuning techniques, including  accurately tuning the partial polarization switching and the band-to-band tunneling drain current, which are not easily realized.
Consequently, asymmetric deviating from the ideal parabolic shape 
result in the generation of undesired harmonics.
In \cite{zhu2023reconfigurable}, a  vertical nanowire ferroelectric tunnel field-effect transistor (ferro-TFET)  that exhibits  a downward-opening parabolic transfer curve was employed as the multiplier device to accomplish  frequency doubling and signal transmission, as depicted in Fig. \ref{fig2}(c)/(d).
However, the fabrication  of nanowire ferro-TFET devices is complex and costly, incompatible with traditional CMOS technology. The parabolic  transfer characteristics of these devices are also facing the challenge of deviating from the symmetrical parabolic shape, leading to the generation of unwanted harmonics.
In this paper, we propose to utilize the standard complementary FeFETs for efficient and economic designs with parabolic transfer characteristic, and further propose feasible and compact reconfigurable frequency multipliers with more multiplications.

\begin{figure*}[t]
\centerline{\includegraphics[width=0.85\textwidth]{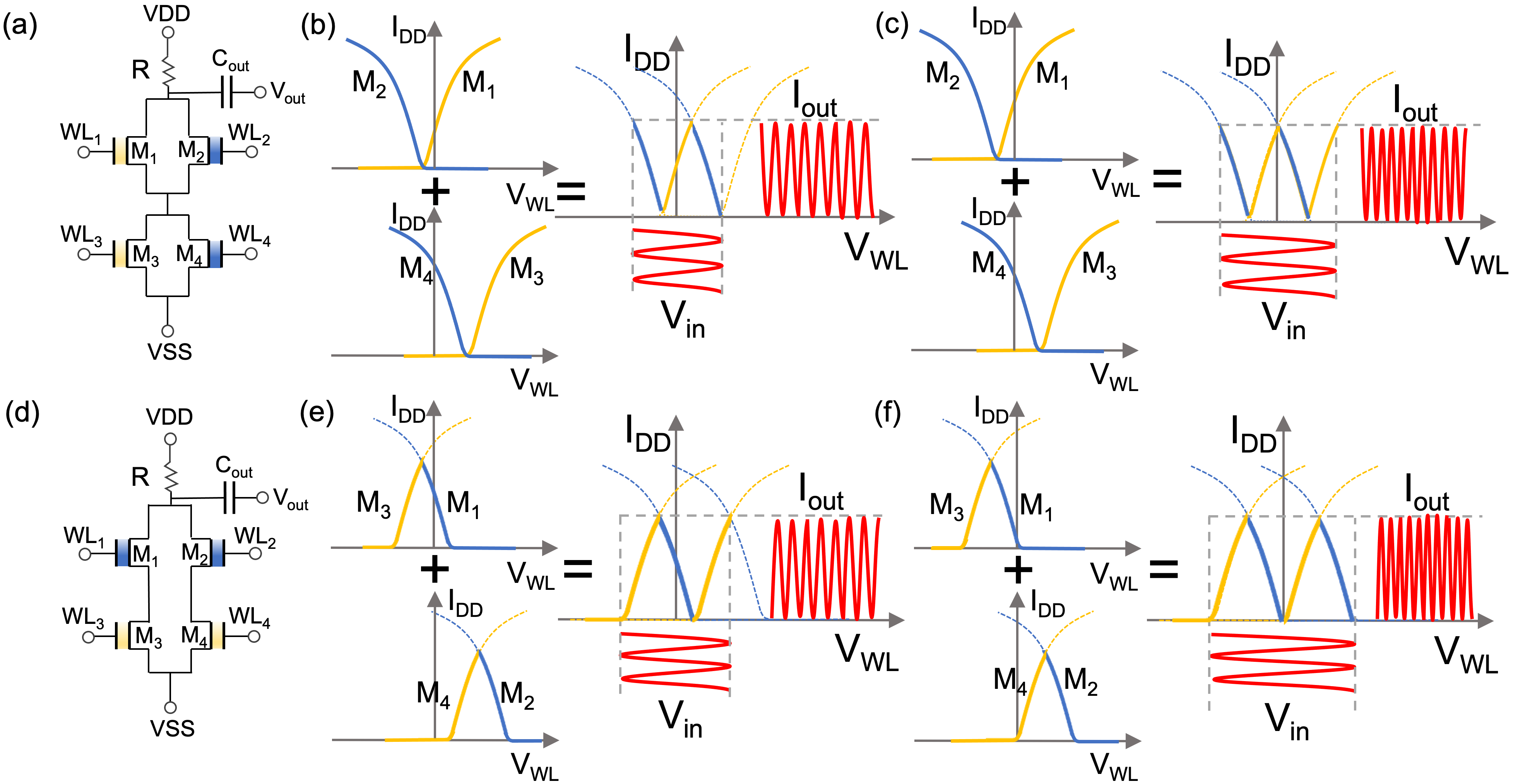}}
\vspace{-1ex}
\caption{
(a) n/p-serial design. (b)/(c) Leveraging the superimposed double-valley shape curve to generate triple/quadruple frequency.
(d) n/p-parallel design. (e)/(f) Leveraging the superimposed double-peak shape curve to generate triple/quadruple frequency.
}
\label{fig5}
\vspace{-3ex}
\end{figure*}

\vspace{-1ex}
\section{Proposed Reconfigurable frequency multipliers}
\label{sec:multipliers}
In this section, we propose four 2FeFET structures implementing reconfigurable frequency multipliers. We  further propose four  reconfigurable multipliers realizing more  harmonics.

\vspace{-1ex}
\subsection{2FeFET structures implementing multipliers}
\vspace{-1ex}
\label{sec:2FeFET}
To address the asymmetrical parabolic characteristic or complex technology issues associated with prior ambipolar FeFET based designs, we propose to use two CMOS-compatible FeFETs based structures as below
to capture  single-parabola-shaped characteristics. 
As depicted in Fig.\ref{fig3}(a),  two nFeFETs are connected in parallel, with their  gate voltages complementary to each other, i.e., $V_{SL}+V_{\overline{SL}}=0$. 
As a result, the  $I_d$-$V_G$ curves of the identical nFeFETs mirror to each other, and the combined transfer characteristic exhibits a symmetrical, upward-opening parabolic shape.
On the contrary, two pFeFETs connected in series with complementary gate voltages as shown in Fig.\ref{fig3}(b) 
yields a symmetrical, downward-opening parabolic transfer characteristic. 
Note that due to the serial connection,  $M_1$ is first programmed to low $V_{TH}$ as a transmission gate to allow programming $M_2$.
Then $M_1$ is programmed.

We also propose 2FeFET structures based on complementary FeFETs. 
As illustrated in Fig.\ref{fig3}(c), nFeFET and pFeFET in parallel 
generates a relatively symmetrical, upward-opening parabolic characteristic. 
This structure  eliminates complementary gate voltages, and can be easily tuned to maintain the parabolic symmetry. On the contrary, as demonstrated in Fig.\ref{fig3}(d), nFeFET and pFeFET in series with the same gate voltage results in a relatively symmetrical, upward-opening parabolic characteristic.
All four 2FeFET structures resemble symmetrical parabolic-shaped characteristic  just like the ambipolar FeFET device discussed in \cite{mulaosmanovic2020reconfigurable, zhu2023reconfigurable}. It can be readily seen that compared with ambipolar FeFETs, our proposed structures are built with CMOS-compatible FeFET devices without complex technology process, and the symmetrical parabolic characteristics can be simply obtained by identical nFeFETs, pFeFETs, or width ratio tuning between nFeFET and pFeFET.  
These structures represent a more general and economic way for the realizations of frequency multiplication. 


The principles of our proposed 2FeFET structures for frequency multiplication are similar to that of ambipolar devices, as shown in  Fig.\ref{fig2}. When employing the 2nFeFETs or complementary FeFETs based structures that exhibit upward-opening parabolic  curves as shown in Fig.\ref{fig3}(a)/(c),  we set the  operating point at the minima of the parabola. When the input signal $V_{in}$ applied at $SL$ transitions from the valley to the peak, the corresponding output current transitions from a peak to a valley, and back to a peak. Consequently, the output signal corresponds to a second harmonic (frequency doubling) of the input signal. Similarly, when employing  the 2nFeFETs or complementary FeFETs based structures exhibiting a downward-opening parabola as shown in Fig.\ref{fig3}(b)/(d), a second harmonic output signal can also be generated. 
As illustrated in Fig.\ref{fig2}(b)/(d), tunable characteristics of FeFETs (see Fig.\ref{fig1}) enable the shift  of the parabolic curves. 
We then place the operating point of input signal at one branch of the curve.
In this configuration, as the input signal transitions from a valley to a peak, the output signal also transitions from a valley to a peak, making the 2FeFET structures signal amplifiers. 

Hereby, our proposed 2FeFET structures can achieve reconfigurable first and second harmonic generation.
Yet, the potential of our proposal has not been fully exploited.
To build general designs that perform scalable frequency multiplications, such as threefold or fourfold, relying solely on the 2FeFET structures with a single parabolic shapes is insufficient. Drawing inspiration from single parabolic curves, we naturally consider superimposing single-parabolic curves  with an  interval to create  double-peak or double-valley shape curves, which can be easily implemented based 
on our proposed structures.

\vspace{-1ex}
\subsection{Reconfigurable multipliers with more multiplications}
\label{sec:4FeFET}
\vspace{-1ex}

We further build four reconfigurable frequency multipliers that support multiple multiplication modes, i.e., first harmonic  (First-H), second harmonic (Second-H), third harmonic (Third-H) and fourth harmonic (Fourth-H), based on previously discussed 2FeFET structures.
As depicted in Fig.\ref{fig4}(a), we  place two parallel connected 2nFeFET structures (Fig.\ref{fig3}(a)) in series. 
The gate voltages of $M_1/M_3$ and $M_2/M_4$ are complementary, i.e., $V_{WL1}+V_{WL2}=0,V_{WL3}+V_{WL4}=0$.  We refer to this structure as the 4n-serial design.
As a result, two upward-opening parabolic curves are superimposed through the “AND” logic, forming a double-valley shape of transfer characteristic as shown in Fig.\ref{fig4}(b)/(c). 
The slight interval between the superimposed upward-opening parabolic curves can be realized by programming the threshold voltages $V_{TH}$ of 4 nFeFETs.
During the write, $M_1$ and $M_2$ are first programmed to low $V_{TH}$ state, passing high voltage to  drains of $M_3$ and $M_4$, respectively. 
The gates of $M_3$ and $M_4$ are then applied with program pulses, followed by programming $M_1$ and $M_2$.


As illustrated in Fig.\ref{fig4}(b),  programming the 4nFeFETs can shift the superimposed double-valley shape curve, such that the amplitude of input signal $V_{in}$ covers a valley and a peak within the double valley, with the midpoint aligning with the operating point of input signal.
As the input signal goes from the valley to the peak, the corresponding output signal transitions between the peak and the valley within the superimposed curve  for three times,
achieving a threefold frequency multiplication, i.e., Third-H. 
The principle for achieving fourfold frequency multiplication is similar,
as illustrated in Fig.\ref{fig4}(c).
The superimposed double-valley curve is programmed such that the amplitude of input signal covers two valleys and one peak, with the midpoint aligning with the operating point of input.
As the input signal transitions from the valley to the peak, the corresponding output transitions between  the peak and the valley within the double-valley curve for four times, generating the fourth harmonic signal.
The  Third-H and Fourth-H generations are determined by the shape of superimposed curve, the   operating point and the amplitude of input.
Similar operations can be applied to other three multiplier designs, and more multiplication modes can be  proposed following above principles.  


As depicted in Fig.\ref{fig4}(d), we build another multiplier by placing two serial connected 2pFeFET structures (Fig.\ref{fig3}(b)) in parallel. 
The gate voltages of $M_1/M_2$ and $M_3/M_4$ are complementary, i.e., $V_{WL1}+V_{WL3}=0, V_{WL2}+V_{WL4}=0$. 
We refer to this structure as the 4p-parallel design. 
Two downward-opening parabolic curves are superimposed through the “OR” logic, forming the double-peak shape transfer characteristic. 
The interval between downward-opening parabolic curves can be realized by programming the $V_{TH}$ of pFeFETs. 
During the write, $M_1$ and $M_2$ are first programmed to low $V_{TH}$ state, passing high voltage to  drains of $M_3$ and $M_4$, respectively. 
The gates of $M_3$ and $M_4$ are then applied with program pulses, followed by programming $M_1$ and $M_2$.
The operation principles for Third-H and Fourth-H generations are similar to that of 4n-serial design, and illustrated in Fig.\ref{fig4}(e)/(f).


The 4n-serial and 4p-parallel designs consume extra circuitry for complementary voltages. 
We hereby propose two multiplier designs to eliminate  complementary voltages.
As depicted in Fig.\ref{fig5}(a), we place two parallel connected nFeFET and pFeFET structures (Fig.\ref{fig3}(c)) in series.  
We refer to this structure as the n/p-serial design. Similar to the 4n-serial design,  this n/p-serial design achieves a double-valley shape curve, and realizes frequency multiplications, i.e., Third-H and Fourth-H generations as illustrated in \ref{fig5}(b)/(c).
Alternatively, as depicted in Fig.\ref{fig5}(d), we place two serial connected nFeFET and pFeFET structures (Fig.\ref{fig3}(d)) in parallel to achieve the double-peak shape curve. We refer to this structure as the n/p-parallel design. Frequency multiplication principles are illustrated in Fig.\ref{fig5}(e)/(f).
\begin{figure}[t]
\centerline{\includegraphics[width=1.0\columnwidth]{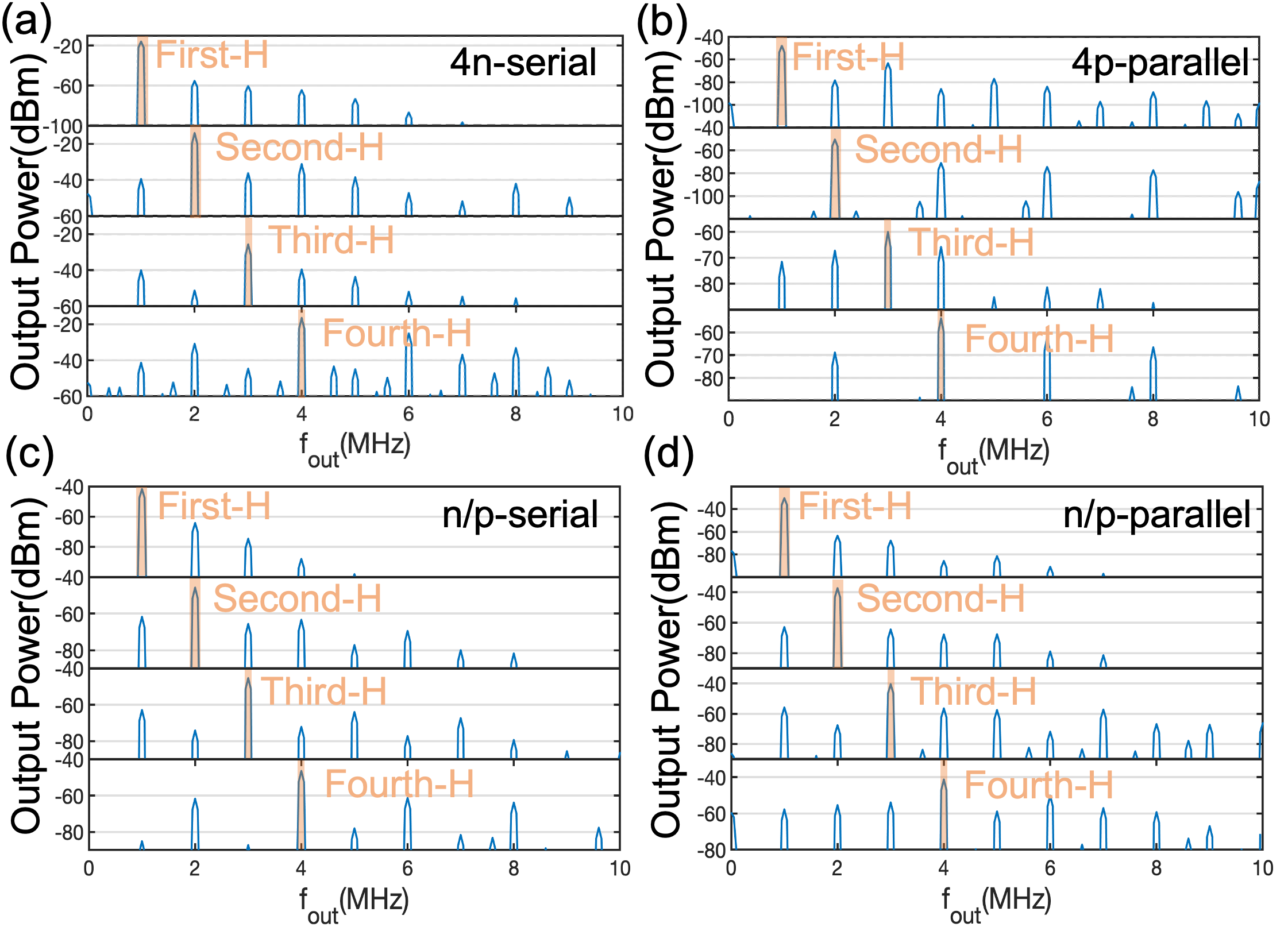}}
\vspace{-1ex}
\caption{
The power spectrum of (a) 4n-serial design, (b)  4p-parallel design, (c)  n/p-serial design, (d) n/p-parallel design, performing frequency multiplications with $f_{in}=1MHz$.
}
\label{fig6}
\vspace{-3ex}
\end{figure}

Overall, we propose above four designs  supporting multiple frequency multiplications, i.e., First-H, Second-H, Third-H and Fourth-H generations, as representatives of scalable frequency multiplier design methodology. 
Based on the 2FeFET structures, we can construct multipliers with more harmonic modes.
\begin{table*}[]
\begin{center}

\caption{Performance of frequency multipliers}
\vspace{-1ex}
\label{tab1}
\begin{tabular}{|c|c|c|c|c|c|c|c|}
\hline
                                          & \textbf{\cite{ref1}} & \textbf{\cite{mulaosmanovic2020reconfigurable}}   & \textbf{\cite{zhu2023reconfigurable}}   & \textbf{4n-serial} & \textbf{4p-parallel} & \textbf{n/p-serial$^\star$} & \textbf{n/p-parallel$^\star$} \\ \hline
\textbf{Sturcture}                        & 5R-11C-8L-3T  & 1R-1C-1T        & 1R-1T           & 1R-1C-4T          & 1R-1C-4T         & 1R-1C-4T               & 1R-1C-4T                     \\ \hline
\textbf{Device}                           & CMOS          & ambipolar FeFET & ambipolar FeFET & FeFET            & FeFET            & FeFET                 & FeFET                        \\ \hline
\textbf{Area($\mu m^2$)} & 280000          & 0.21        & 0.01        & 0.0256           &0.0448            & 0.182              & 0.190                    \\ \hline
\textbf{Max Frequency(MHz)}                & 20000      & 1        & 0.01        & 5000         & 10000      & 100              & 1000                     \\ \hline
\textbf{Power($\mu$W)}                        & 20000     & 4.8        & 0.00151        & 3.35        & 3.12         & 8.05              & 9.34                    \\ \hline
\textbf{Supply Voltage(V)}                & 2.6           & 3.5             & 0.5             & 1                & 1                & 1                     & 1                            \\ \hline
\textbf{Multiplication}                       & x2            & x1,x2           & x1,x2           & x1,x2,x3,x4      & x1,x2,x3,x4      & x1,x2,x3,x4           & x1,x2,x3,x4                  \\ \hline
\textbf{Reconfigurablility}                   & no            & yes             & yes             & yes              & yes              & yes                   & yes                          \\ \hline

\end{tabular}

 \begin{flushleft}
 \scriptsize
$\star$: Eliminating the complementary voltage generator circuitry.\\
 \end{flushleft}
\vspace{-4ex}

\end{center}
\vspace{-3ex}
\end{table*}


\vspace{-1ex}
\section{Validations and Evaluations}
\label{sec:eval}
\vspace{-1ex}
In this section, we validate the proposed frequency multipliers with various multiplication modes. We then evaluate and compare the designs with prior works. 
Finally, a case study of reconfigurable multiplier is demonstrated. 45nm Preisach FeFET model \cite{ni2018circuit} is utilized in the Cadence simulations. The size of pFeFET is tuned to ensure that the conducting current is nearly the same as that of nFeFET.

\vspace{-1ex}
\subsection{Frequency multiplier validations and comparisons}
\label{sec:validate}
\vspace{-1ex}
We validate the frequency multiplication functions of our proposed reconfigurable multipliers. 
Fig.\ref{fig6} presents the output power spectrum  of the four proposed designs in different frequency multiplication modes. 
It can be observed that all four designs confirm their frequency multiplication functionality. 
Although non-ideal transfer characteristics inevitably lead to extra harmonics, the output power of  target harmonic frequency  is at least 10dBm higher compared to other harmonic components. 

\begin{figure}[t]
\centerline{\includegraphics[width=0.9\linewidth]{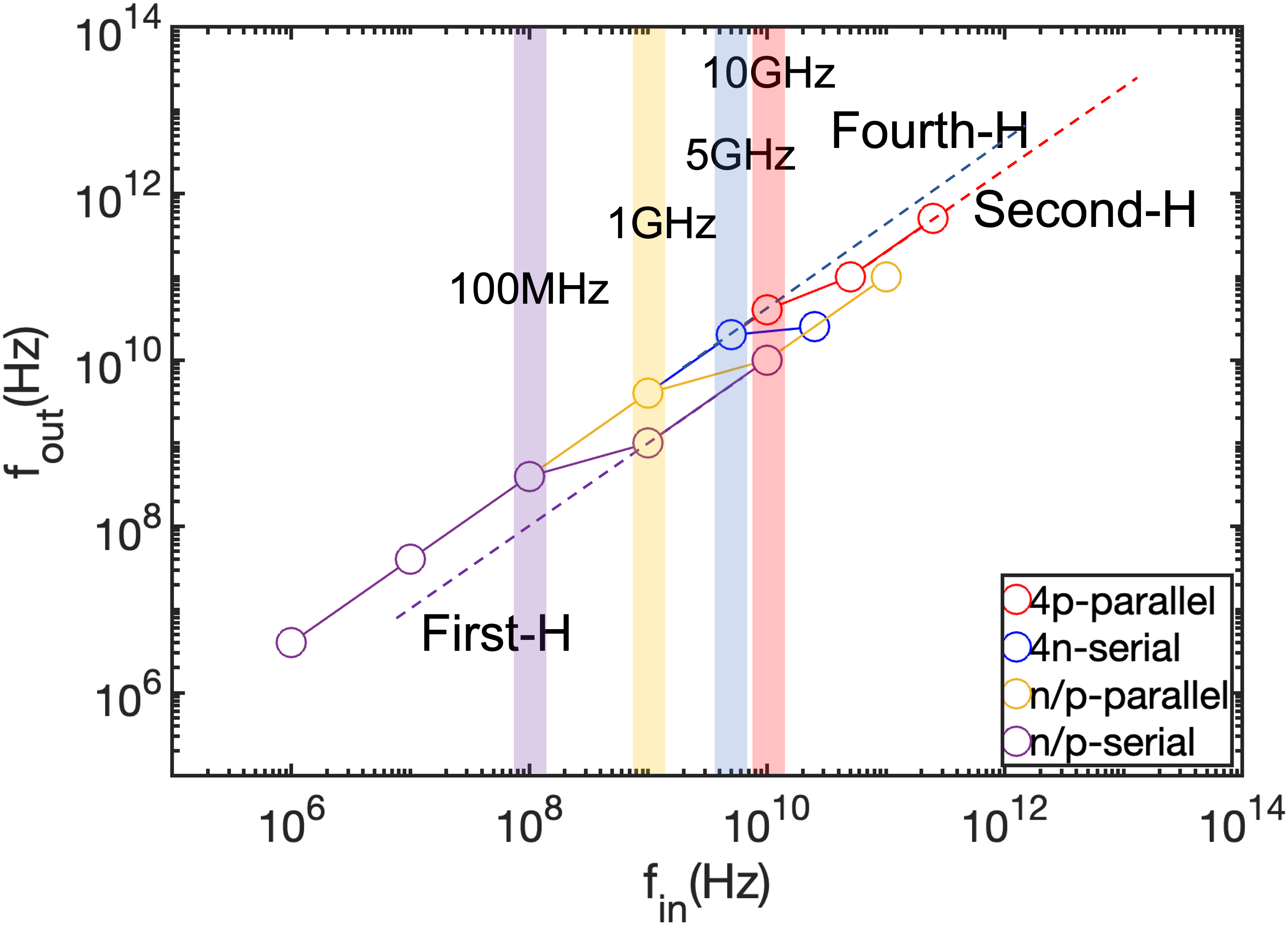}}
\vspace{-1ex}
\caption{
Maximum operating frequencies of  multiplier designs.
}
\label{fig7}
\vspace{-3ex}
\end{figure}

The maximum operating frequency represents the capability of multipliers. 
Taking frequency quadrupling as an example. 
As shown in Fig.\ref{fig7}, for the 4n-serial design, when the input signal frequency exceeds 5GHz,  the Fourth-H component starts to decline,  and the First-H dominates by the time it reaches 10GHz. 
The 4p-parallel design  functions well  even at 10GHz of input signal, but when the input frequency  reaches 100GHz, the Second-H starts to dominate. 
As for the n/p-serial design, its maximum operating frequency is 0.1GHz, while for the n/p-parallel structure, the maximum operating frequency is 1GHz.

Table \ref{tab1} summarizes the performance metrics of the proposed and existing frequency multipliers. It is evident that, compared to traditional CMOS multipliers\cite{ref1}, the reconfigurable multipliers based on FeFETs offer the advantages of smaller area and extremely low power consumption. 
However, the maximum operating frequencies of the  reconfigurable multipliers in \cite{mulaosmanovic2020reconfigurable,zhu2023reconfigurable} are much lower than those of traditional multipliers. 
Our proposed four reconfigurable multipliers   have much higher maximum operating frequencies than those in \cite{mulaosmanovic2020reconfigurable,zhu2023reconfigurable}. Furthermore, the maximum operating frequency of the 4p-parallel design  approaches that of traditional CMOS multipliers.
All four proposed reconfigurable multipliers  offer four different frequency multiplication modes (and potentially more modes given the scalability), surpassing traditional multipliers with fixed multiplication mode \cite{ref1} and the two modes in \cite{mulaosmanovic2020reconfigurable,zhu2023reconfigurable}.

\begin{figure}[t]
\centerline{\includegraphics[width=0.75\columnwidth]{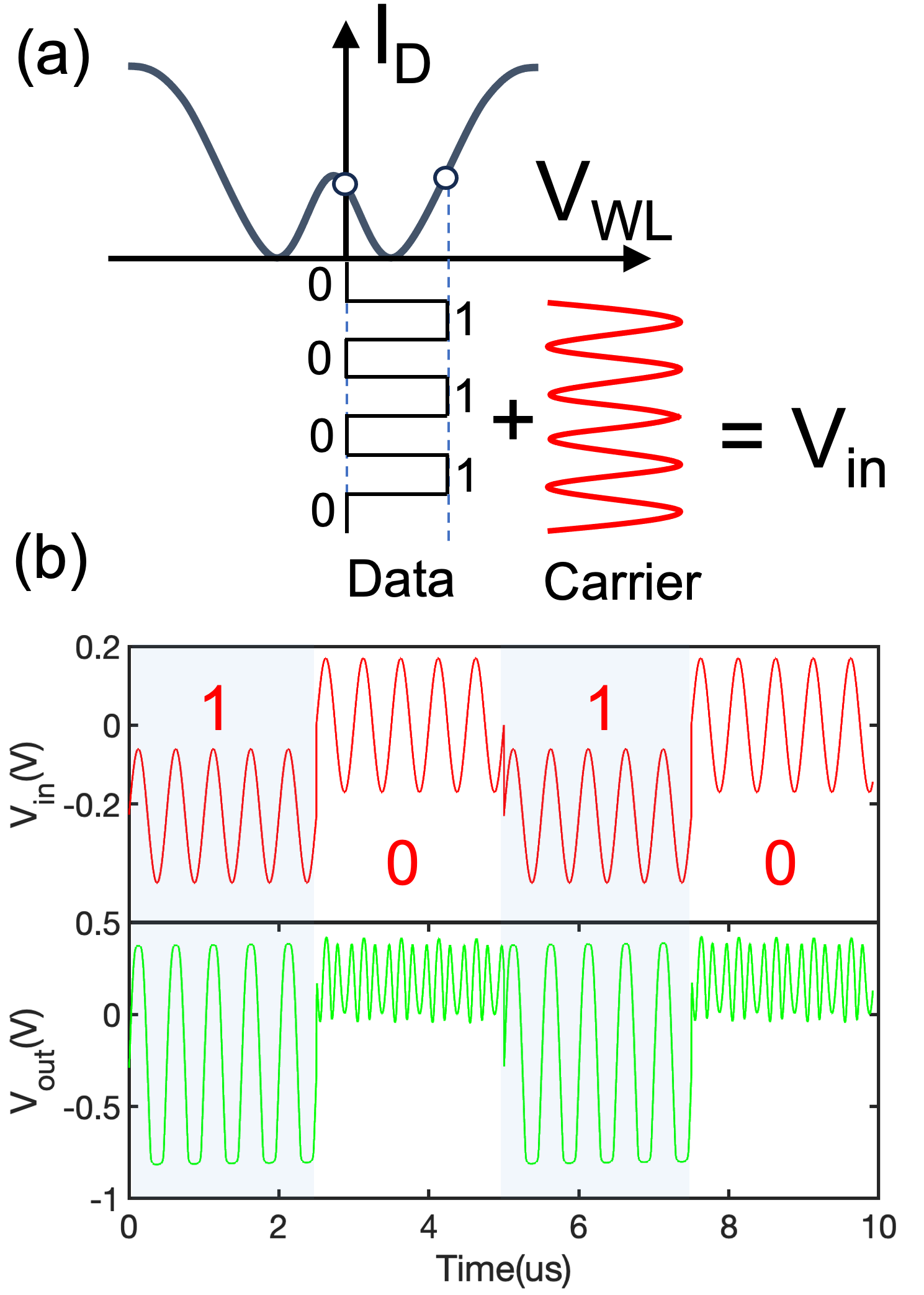}}
\vspace{-1ex}
\caption{
(a) Illustration of  FSK implementation, with  input signal $V_{in} = V_{data} + V_{carrier}$. 
(b) Corresponding output signal.
}
\label{fig8}
\vspace{-4ex}
\end{figure}

\vspace{-2ex}
\subsection{Case study: Frequency Shift Keying}
\vspace{-1ex}
We showcase that
our  proposed frequency multiplier designs  can achieve frequency shift keying (FSK) functionality, which is widely used in communication systems \cite{watson1980fsk}. 
The FSK can be implemented by applying the data pulses and carrier signals at the input of our multipliers, i.e.,  $V_{in} = V_{data} + V_{carrier}$. 
The valley and the peak of data pulses are aligned with the operating points of the multipliers 
as depicted in Fig.\ref{fig8}(a), respectively.
When the data signal is at '1' level, the input signal oscillates at the operating point that achieves First-H, i.e., signal transmission.
When the data signal is at '0' level, aligning with the operating point that achieves Third-H, the output frequency is triple of input.
In this way, the output frequency periodically shifts between single frequency and triple frequency, as shown in Fig.\ref{fig8}(b). 
As can be seen, the alignment of data pulses with the operating points within the transfer curve of multipliers is the key enabler of FSK. 
This alignment can be realized by   adjusting the peak/valley amplitude of data pulses and carrier signal, while maintaining the FeFET programming configurations.
However, this method can only achieve the frequency shifts between single  and double/triple/quadruple frequencies, as well as the shift between double and quadruple frequencies\footnote{Aligning the valley/peak of data pulses with the operating points  corresponding to Second-H (Fig.\ref{fig2}(a)/(c))/Fourth-H (Fig.\ref{fig4}(c)/(f)) and (Fig.\ref{fig5}(c)/(f)).}.
Altering the FeFET programming configuration of multipliers, on the contrary, can generally achieve the shifts between any frequency modes, while consuming extra programming time.

\vspace{-2ex}
\section{Conclusion}
\label{sec:conclusion}
\vspace{-1ex}
We exploit the programmable characteristics of complementary FeFETs to build  four 2FeFET structures  for reconfigurable frequency doubling and signal transmission functions.
These 2FeFET structures resemble similar parabolic-shape curves to prior ambipolar FeFETs, while requiring much simpler technology and characteristic tuning.
Based on 2FeFET structures,  four novel reconfigurable frequency multipliers allowing for seamless switching between frequency multiplication modes are proposed.
The performance of the propsoed multipliers are evaluated, and an example case of FSK is demonstrated.
As emerging FeFETs continue to evolve, this work will provide insights to more compact, efficient and flexible frequency multiplier designs with enhanced multiplication modes compared to traditional counterparts and existing designs.
\vspace{-1ex}
\section*{Acknowledgements}
\vspace{-1ex}
This work was supported in part by  NSFC (92164203, 62104213), Zhejiang Provincial Natural Science Foundation (LD21F040003, LQ21F040006),  National Key R\&D Program of China (2022YFB4400300).

\bibliographystyle{ieeetr}
\vspace{-1ex}
\bibliography{bib}


\end{document}